# Towards Asimov's Psychohistory:
# Harnessing Topological Data Analysis, Artificial Intelligence and Social Media data to Forecast Societal Trends

Isabela ROCHA[1]
Institute of Political Science
of the University of Brasília

**Abstract:** In the age of big data and advanced computational methods, the prediction of large-scale social behaviors, reminiscent of Isaac Asimov's fictional science of Psychohistory, is becoming increasingly feasible. This paper consists of a theoretical exploration of the integration of computational power and mathematical frameworks, particularly through Topological Data Analysis (TDA) (Carlsson, Vejdemo-Johansson, 2022) and Artificial Intelligence (AI), to forecast societal trends through social media data analysis. By examining social media as a reflective surface of collective human behavior through the systematic behaviorist approach (Glenn, et. Al, 2016), I argue that these tools provide unprecedented clarity into the dynamics of large communities. This study dialogues with Asimov's work, drawing parallels between his visionary concepts and contemporary methodologies, illustrating how modern computational techniques can uncover patterns and predict shifts in social behavior, contributing to the emerging field of digital sociology – or even, Psychohistory itself.

---

[1] Master of Science and PhD Candidate at the Institute of Political Science of the University of Brasília (IPOL-UnB). Coordinator of the Working Group on Strategy, Data and Sovereignty of the Study and Research Group on International Security of the Institute of Foreign Affairs of the University of Brasília (GEPSI IREL-UnB). Member of the Public Information and Elections Research Group at the Institute of Political Science of the University of Brasília (IPê UnB) and the International Political Science Association 50th Research Committee on the Politics of Language (IPSA RC50).

**Introduction**

In Isaac Asimov's Foundation series (1982), Psychohistory is a fictional scientific discipline that combines history, sociology, and statistical mathematics to predict the future behavior of very large groups of people. Asimov's concept relies on the idea that while the actions of an individual are unpredictable, the behavior of a large group can be forecasted with a high degree of accuracy. This science was developed by the character Hari Seldon, who uses it to foresee the fall of his futuristic society and to establish the Foundation to shorten the ensuing period of chaos and rebuild a new, stable civilization after the inevitable collapse of the Galactic Empire.

The advent of big data and sophisticated computational techniques has transformed our capacity to analyze and predict complex social phenomena (Vazquez, 2022), drawing reality ever closer to the realm of fiction. And as science fiction has often served as a precursor to scientific advancements, providing visionary frameworks that inspire real-world innovations, what was once a speculative idea is now inching toward reality as our analytical capabilities evolve. In this context, this article delves into the intersection of modern computational power, societal data availability and mathematical frameworks, particularly Topological Data Analysis (TDA) and Artificial Intelligence (AI), to suggest early-stage models theoretically capable of forecasting societal trends, in hopes of actualizing Asimov's vision of Psychohistory into reality. In short, this study hypothesizes that by leveraging TDA and AI on social media data, it is possible to predict societal trends with significant accuracy, akin to Asimov's Psychohistory.

This fictional scientific discipline becomes feasible as, in recent years, scientists have witnessed an explosion in the availability of social media data, offering extremely large datasets containing a plethora of samples of human interactions and behaviors – possibly enabling Asimov's axiom that the population whose behavior was modelled should be sufficiently large to represent the entire society. These platforms provide a reflective surface of collective human activity, enabling researchers to discern patterns and identify shifts in political or societal dynamics (Sunstein, 2017, Sanches, Silva, 2023, Coleman, 2012, Bennet, Segerberg, 2012), and now that computational power and ingenuity has increased, either by the advancement of hardware or AI, one may consider that large social media derived data might be used for feasible prediction of outcomes (Rhukuzage, 2020, Vazquez, 2022). In fact, it seems now, that rather than the issue of limited available data, scientists' challenges involve discovering powerful enough hardware, or sufficiently ingenious programming, to process such sheer amount of information.

Regardless of the inherent potential that comes from scientists' ease of access to social media data, the integration of these computational methods into the realm of social prediction remains still in its nascent stages – specially as the use of AI, for example, is in its budding stages when it comes to the use of such tools by those in the not so hard sciences – popular models such as ChatGPT have, after all, only just been popularized. Thus, this study builds on the legacy of Asimov's Psychohistory, drawing parallels between his visionary concepts and contemporary methodologies in hopes of inspiring researchers to holistically incorporate the studies of history, psychology, mathematics, and computation by inspiring the design of methodologies capable of predicting political and/or societal outcomes.

The objectives of this exploratory research are multifaceted, but can be broken into two phases. The first step involves a comprehensive review of the use of social media analysis within the social sciences, aiming to understand how these platforms provide insights into psychological, political, and societal trends. The second step focuses on demonstrating the applicability of TDA in identifying significant patterns within social media data. This step serves as the foundation for a later integration of AI to enhance the predictive accuracy of these patterns, moving towards the development of mathematical models capable of forecasting societal outcomes. Finally, the study will discuss the broader implications of these findings for the field of the Computational Social Sciences and the potential evolution towards studies inspired by Psychohistory, integrating advanced models of large dataset analysis.

TDA is particularly useful in this context due to its ability to reveal the underlying structure of complex, high-dimensional data. According to Gunnar Carlsson (2022), TDA focuses on identifying and analyzing topological features such as connected components, loops, and voids, which persist across multiple scales of data resolution – all useful for the understanding of social phenomena expressed in datasets. The inherent difficulty in understanding large datasets, especially in Social Media Research, arises from the sheer volume and complexity of the data involved. Even modest studies can involve hundreds of thousands of pieces of content and accompanying metadata. This complexity is compounded by the varied and intricate sources of the datasets used for analysis and the construction of predictive models. These challenges highlight the necessity of employing TDA, as its capabilities are uniquely suited to manage and elucidate the complexities inherent in such large-scale data.

In other words, TDA's multi-scale approach is critical for understanding the intrinsic geometry and topology of data, especially when dealing with the vast datasets generated by social media. Additionally, social media data inherently comes from network structures, consisting of nodes (users) and edges (interactions). These networks are often characterized by complex relationships and high-dimensional interactions, which are challenging to analyze using traditional methods. TDA, however, excels at uncovering patterns that traditional statistical methods might miss, and through tools like Persistent Homology, one may capture the essential features of data, even in the presence of noise, by providing a robust summary of the data's shape. This capability is invaluable for analyzing social media data, where patterns of human behavior and interaction are often subtle and distributed across various scales.

By applying TDA to social media data, researchers can identify significant patterns and structures that reflect underlying social or political behaviors and dynamics (Rocha, 2023, Rocha, 2024, Carlsson, Vejdemo-Johansson, 2022). These patterns, once identified, can be further analyzed and modeled using AI to enhance their predictive accuracy. This combination of TDA and AI enables the development of sophisticated mathematical models that can forecast societal outcomes with a higher degree of precision. Thus, TDA not only strengthens the analytical foundation of this research but also bridges the gap between raw data and actionable insights, paving the way for advancements in computational social sciences and the realization of predictive models inspired by the concept of Psychohistory.

Through this comprehensive literature review, we seek to explore the initial stages of realizing Psychohistory as a legitimate field of study. By examining the application of TDA and AI, we aim to understand how these tools can uncover hidden patterns within social media data which is inherently social and human in nature, contributing to the prediction of large-scale social behaviors. This exploration not only seeks to bridge science fiction and real-world applications but also contributes to the emerging field of digital sociology – or even, Psychohistory itself.

This paper is structured as follows: It begins with a review of the relevant literature on social media analysis and its applications in some of the social sciences, particularly sociology, political science and psychology. Following this, the methodological approach is presented, detailing the data sources, analytical techniques, and the feasibility of designing future predictive models using TDA and AI. The subsequent section discusses the findings, highlighting key patterns and their implications, as well as a workflow proposition to enable further research and a few pilot models. The paper concludes with a discussion of the broader impacts of this research and potential future directions for the field.

**Literature Review**

The Ruliad, a term coined by Stephen Wolfram (2023), describes the abstract and intertwined totality of all possible computational trajectories or processes that can occur within a Technosphere. It represents the infinite possibilities of computational and procedural operations, encapsulating every conceivable configuration of computational systems. Through computational language, such as the Wolfram Language, knowledge and data across diverse fields – from physics and mathematics to biology, economics, and political science – can be expressed and operationalized. In other words, the Ruliad serves as a theoretical space where any concept, theory, or paradigm can be formalized and explored computationally. By leveraging the computational power and mathematical frameworks represented within the infinitude of the Ruliad, particularly through TDA and AI, the uncovering of hidden patterns within social media data that reflects large-scale of social behavior becomes feasible.

In short, the very definition of Wolfram's Ruliad designs its capability to accommodate and process the vast complexities of social phenomena by implying that predictive models for societal trends, which were once considered speculative, can now be systematically developed and validated as the Ruliad's framework allows for the exploration of various computational pathways, enabling researchers to simulate and analyze different scenarios and their potential impacts on societal dynamics. And this newfound possibility is the backbone of this exploratory research.

TDA allows for the Ruliad's exploration by providing a robust framework for analyzing and understanding the vast array of possible configurations within complex datasets. The Ruliad, as described by Stephen Wolfram, encapsulates the infinite possibilities of computational trajectories and processes, representing every conceivable configuration of computational systems, and, TDA, with its focus on identifying and analyzing topological features such as connected components, loops, and voids, enables researchers to systematically explore these configurations. This suggests that, by applying TDA, researchers can test and validate any model within the Ruliad, making it possible to uncover hidden patterns and relationships that inform our understanding of large-scale social behaviors. This capability is particularly crucial for realizing Psychohistory, as it allows for the comprehensive exploration of the myriad ways in which individual interactions and behaviors can aggregate to form societal trends. In other words, TDA provides the necessary tools to map out these interactions across multiple scales, ensuring that even the most subtle and complex patterns are captured and then analyzed.

Furthermore, TDA's ability to handle high-dimensional and noisy data makes it ideal for navigating the complexities of social media datasets. By filtering out irrelevant information and highlighting significant topological features, TDA facilitates a deeper understanding of how various computational pathways can influence societal dynamics. This approach not only enhances the development of predictive models but also ensures that these models are grounded in a thorough and systematic exploration of the Ruliad.

In the following subsection we shall investigate aspects necessary to the realization of Psychohistory by outlining the necessary toolkit employed for a later research design: Social Media analysis, the Behaviorist approach and how TDA might be used in the scope of this overarching research.

*Social Media as a looking glass*

The explosion of social media platforms over the past two decades has fundamentally transformed the landscape of human interaction and societal dynamics. These platforms serve as a looking glass through which collective human behavior can be observed, analyzed, and understood as the vast amounts of data generated by social media provide an unprecedented window into the nuances of human interaction. Stephen Coleman (2005) was one of the early scholars to discuss the reconfiguration of political representation in the digital age as digital platforms enable direct and immediate interactions between representatives and constituents, reshaping the traditional mechanisms of democratic engagement. This new mediation not only enhances the responsiveness of political systems but also reflects the evolving nature of public discourse and participation in the digital era. Bennett and Segerberg (2012) further advanced the understanding of social media's impact by introducing the concept of Connective Action as Digital Social Media facilitates personalized forms of political and social mobilization. Unlike traditional collective action, which relies on formal organizations, connective action thrives on the decentralized and personalized nature of digital communication.

This connective action, which could be understood as a shift in the way humans relate to each other through various digital environments, showcases a profound reflection between the dynamics of social media and the real world as Social Media platforms do not merely mirror offline interactions but actively shape and influence them, creating a feedback loop where digital behaviors and real-world actions continuously interact. The personalized and decentralized nature of connective action on social media means that movements and trends emerging online can quickly manifest in tangible societal changes. This interplay is evidenced by the rapid mobilization of social movements, the spread of information (and misinformation), and the formation of echo chambers that reinforce existing beliefs and attitudes (Sunstein, 2017). This interaction between online and offline behaviors underscores the critical role of social media in contemporary political and social dynamics, reinforcing the importance of studying these platforms through advanced computational methods such as TDA and AI to capture the complexities of modern human interaction.

This shift, characterized by the connective action, underscores the role of social media as a dynamic mirror of societal engagement, where individual actions and interactions coalesce into broader social movements. In this sense, Cass Sunstein (2017) examined the dual role of social media as both reflective and formative forces in democracy by noting that social media algorithms often create echo chambers that reinforce existing biases leading to greater political polarization – a phenomena identified in several democracies across the globe, and observed in the scope of my work (2023, 2024a, 2024b) by shapes that could be potentially described through different gaussian-function configurations. In this sense, Sanches and Silva (2023) also addressed the implications of digital technologies for social participation and political engagement by highlighting the potential of digital tools to enhance civic participation, while also noting the challenges of ensuring inclusive and accessible digital governance.

Finally, Rhukuzage (2020) explored the transformative impact of social media on sociological research, delving into the distinctions between digital sociology and the sociology of the digital. He argued that the integration of digital tools and methodologies is essential for capturing the complexities of contemporary social dynamics. In this sense, Social Media, as a looking glass, offers a critical vantage point for understanding how digital phenomena influence social structures and behaviors. Vazquez (2022) also emphasized the methodological challenges and opportunities in modeling social media phenomena by highlighting the importance of integrating computational tools and big data analytics to uncover hidden patterns within social media data.

Therefore, it is worth reviewing a few concepts stemming from the behaviorist sciences as they dabble into cultural research – particularly as the concepts of Metacontingency, Culturo-Behavioral Lineage and Cultural Transmission, Macrobehavior, Macrocontingency and Cultural Cusps (Glenn et. Al, 2016). Metacontingency refers to the interrelated contingencies that produce aggregate outcomes, such as societal trends or collective behaviors, influenced by environmental selecting conditions. In the context of social media, Metacontingencies can explain how certain behaviors and ideas gain prominence and reinforcement within online communities. This concept involves the transmission of operant behaviors through individual repertoires over time, forming a lineage of cultural practices. On platforms like Twitter, behaviors such as tweeting and retweeting create a lineage of cultural norms and values shared within a community (Rocha, 2024). Traditionally, cultural transmission is the process through which behaviors, practices, and beliefs are passed from one generation to another.

Macrobehavior refers to socially learned operant behaviors common within a group or society, reinforced by consistent operant contingencies. On social media, Macrobehaviors manifest as persistent interaction patterns, such as echo chambers, where users engage in repetitive behaviors that reinforce group cohesion and polarization. As for Macrocontingency, this concept describes how individual or group behaviors can lead to significant societal changes, representing the cumulative effect of many operant behaviors. In social media, Macrocontingencies illustrate how collective behaviors, even when driven by diverse contingencies, accumulate to produce substantial social impacts. Finally, a Cultural Cusp represents a confluence of unique behavioral contingencies that result in significant sociocultural changes.

These designations are, by their own definition, useful for employment in advanced methodologies – such as TDA. While there have been a few works addressing the use of TDA for psychological studies, such as: autism detection, by employing a feature extraction technique based on TDA in order to classify autistic subjects from typically developing ones during a visual cognitive task (Majumder et al., 2020); well-being, by investigating remote work travel experiences using topological analysis on Instagram posts to reveal distinct elements that influence well-being (Chevtaeva et. Al, 2023); and the understanding of psychosis, by exploring methodological expansions in psychometrics to evaluate the topological structure of psychotic experiences and assess longitudinal changes (Argolo et. Al, 2020), there have been few studies seeking to illuminate Macrobehaviors, Macrocontingencies, and Cultural Cusps in large social-media derived datasets as well as when or how psychological and political trends could be predicted through TDA.

In this sense, the behaviorist approach, connected to TDA-powered computational political analysis, might be instrumental in realizing the concept of Psychohistory, as, by leveraging the systematic and pragmatic nature of behaviorism, researchers can develop more accurate models for predicting political and societal trends. For example, Macrobehavior, or the persistent interaction patterns within a group, can be analyzed with TDA to identify stable topological features that indicate cohesive group behavior. This can be particularly useful in understanding the dynamics of political factions and predicting their long-term stability and responses to external events. Macrocontingencies, the cumulative effects of individual or group behaviors that lead to significant societal impacts could also be studied through TDA as, by mapping these identified Macrocontingencies as nodes inside an inherent network, researchers can predict large-scale political and social changes, such as shifts in public opinion or the emergence of new political movements. Finally, as Cultural Cusps represent critical points where unique behavioral contingencies converge to create significant sociocultural changes, TDA could help identify these pivotal moments by highlighting the interactions that lead to transformative political and social phenomena.

This all could potentially allow for the anticipation of major shifts in the political landscape. By integrating these behaviorist concepts with TDA, researchers can develop robust predictive models that capture the complexity of social media interactions and their broader societal impacts. This approach not only enriches the theoretical understanding of digital sociology but also provides practical tools for analyzing and anticipating complex social dynamics, bringing us closer to the realization of Psychohistory.

**Methodological approaches**

*The road so far*

By utilizing TDA to identify significant topological features in social media data and integrating AI to enhance the predictive accuracy of these patterns, this overarching research aims to build robust mathematical models. These models can potentially predict societal outcomes with a high degree of precision, aligning with Asimov's vision of Psychohistory. Thus, the previously defined concept of the Ruliad not only enriches the theoretical underpinnings of this study but also lays a concrete foundation for the practical application of computational methods in forecasting social behavior. This intersection of advanced computational theories and practical data analysis techniques marks a significant step forward in the field of the Computational Social Sciences, opening new avenues for research and application.

As previously stated, social media data will be inherently modeled through networks, however researchers must now stretch their perspective towards TDA as to enable more refined predictive or analysis models. Serrano, Serrano, and Gómez's work (2020) offers profound arguments in favor of this shift as their study critiques the limitations of standard graph-based methods that only account for pairwise interactions, arguing for the inclusion of higher-order interactions inherent in real-world networks – which also happen to be present in social media networks. The authors propose an advanced mathematical framework using simplicial complexes to capture these multi-agent interactions, which cannot be adequately represented by conventional network theory.

This is because simplicial complexes allow for the examination of structures where interactions among three or more agents are common, an approach that provides a richer and more nuanced understanding of connectivity in complex networks. The introduction of multi-parameter boundary and coboundary operators facilitates the computation of higher-order degrees of adjacency, enhancing our ability to analyze the structural properties of networks across various domains. This methodological advancement is particularly relevant to the study of social media data, where the intricate web of interactions extends beyond simple pairwise relationships. In other words, by employing the concept of higher-order degrees, such as maximal upper simplicial degree and general adjacency degree, it is possible to uncover hidden patterns of collaboration and influence within social media networks. This approach aligns seamlessly with my research, where TDA is utilized to reveal the underlying structures of social media interactions, offering a more comprehensive view of societal dynamics.

Moreover, the work of Serrano et al. underscores the importance of distinguishing between different types of simplicial adjacency, such as lower and upper adjacency, to accurately model the complexity of social interactions. This distinction is crucial for developing predictive models that can account for the multi-faceted nature of human behavior and social influence, ultimately enhancing the precision of our forecasts. Integrating these advanced topological methods with Machine Learning techniques not only improves our analytical capabilities but also addresses the inherent challenges of processing and interpreting vast amounts of high-dimensional data.

By leveraging the robust framework provided by TDA and the innovative techniques proposed by Serrano and colleagues, we can move closer to actualizing the vision of Psychohistory, where the prediction of large-scale social behaviors becomes a feasible scientific endeavor. In this context, Arfi's (2024) groundbreaking work on the use of Persistent Homology, Machine Learning, and deep neural networks in topological data analysis of democracy survival provides a compelling example of how these advanced methodologies can be applied to social sciences – particularly, political science. Arfi's research employs TDA to uncover topological invariants within high-dimensional datasets, using Persistent Homology to identify features such as one-dimensional loops that persist across scales. By integrating these topological features into Bayesian survival analysis models through functional principal component analysis (FPCA), Arfi demonstrates their significant impact on explaining variance in survival data. His empirical analysis of democracy survival from 1950 to 2010, which includes multi-frailty survival analysis to account for recurrent democratic breakdowns, reveals that topological features are as critical as traditional covariates like civil society and party institutionalization, aligning with my findings regarding the use of social media data analysis as means for understanding societal trends such as political polarization and political personalism (Rocha, 2023, 2024).

*Topological Data Analysis pilots*

Arfi's integration of TDA with Machine Learning and Bayesian approaches provides a robust framework for analyzing high-dimensional data, highlighting the global topological structures that traditional methods might overlook. This methodology might be employed in the scope of the systematic behaviorist approach here proposed, which leverages TDA to uncover underlying patterns within social media data, thereby enhancing possible predictive models of social behavior. In this subsection, we shall investigate how the behaviorist perspective might be understood in previously collected data, leading us to the proposition of an actual workflow.

The process begins with automated data extraction from various social media platforms, utilizing web scraping tools and APIs. This step ensures the collection of vast datasets encompassing diverse types of content interactions, such as posts, comments, likes, and shares, as well as other significant metadata. Following data collection, the next phase involves applying TDA to the network structures inherent in social media data. By representing interactions as networks with nodes (users) and edges (interactions), TDA is employed to extract significant topological features. Toolkits have been developed for Social Media data scraping, from Twitter (Rocha, 2023, 2024a, 2024b), to Telegram (Rocha, Silva, Vicentini, 2024) – and will continue to be developed as to enable this research. Next steps include developing toolkits for various Social Media networks.

Next, these topological features serve as the foundation for identifying the previously discussed behaviorist trends, particularly the Metacontingencies, Culturo-Behavioral Lineages and Cultural Transmissions, Macrobehaviors, Macrocontingencies, and Cultural Cusps. The integration of AI models trained by specialized professionals in this phase is crucial to enhance the predictive power and analytical depth:

Macrobehavior refers to socially learned operant behaviors common within a group or society, reinforced by consistent operant contingencies. TDA is utilized to identify stable topological features that represent cohesive group behavior – and, in this sense, the identification of Nuclear Constellations Persistent Homologies (Rocha, 2024), might be of use as these indicate groups of similarly-minded interactions, or alternatively political personalism. The identification of these structures might provide a basis for Machine Learning models to predict changes in Macrobehaviors.

The approach to identifying Metacontingency could potentially involve the use of TDA to pinpoint persistent topological features in the network that correspond to collective behavioral patterns. These features – and the content present in these nodes – would then fed into machine learning algorithms, which are then trained to recognize these patterns and predict their evolution over time. To analyze culturo-behavioral lineage and cultural transmission, NLP techniques could be potentially applied to the textual content to detect patterns indicative of cultural transmission – but of course, heavily supervised by researchers. TDA would then be used to map the evolution of these behaviors within the network. In this sense, Recurrent Neural Networks (RNNs) and Long Short-Term Memory networks (LSTMs) could also be suitable for modeling the temporal dynamics of cultural transmission in social media interactions, capturing how these behaviors propagate and evolve over time, but this would require another particular research altogether.

As for the Macrocontingencies, TDA might be employed to map these cumulative behaviors as nodes within the network. Machine learning models could be used to predict large-scale societal impacts resulting from these Macrocontingencies, and Graph Neural Networks (GNNs) would be particularly well-suited for processing and analyzing the network-based representations of Macrocontingencies, allowing for a detailed understanding of how individual behaviors aggregate to produce substantial social changes. Finally, as Cultural Cusps represent critical points where unique behavioral contingencies converge to create significant sociocultural changes, TDA might used to highlight interactions that lead to transformative political and social phenomena, identifying these pivotal moments within the data, as well as the actors or the specific content that culminates in the consolidation of a Cultural Cusp.

In this sense, it is wise to consider the role played by network gatekeepers, as they play a crucial role in directing the flow of information and attention within social media networks (Garimella, et. Al, 2018). By strategically filtering and amplifying certain narratives, gatekeepers influence the formation and intensification of polarization. Gatekeepers, by selectively curating content, can accelerate the development of Macrocontingencies and Cultural Cusps by emphasizing divisive or emotionally charged topics. Their role in reinforcing echo chambers and directing attention towards specific narratives amplifies the impact of these phenomena. Integrating AI into the study of gatekeepers involves using machine learning models to identify and track these influential actors and their activities. By mapping the network positions and influence patterns of gatekeepers, TDA and AI can uncover the pathways through which attention is redistributed and polarization is reinforced.

Machine learning models, by incorporating the identification of gatekeepers and their content, can then be trained to predict the occurrence of these cultural cusps. Hybrid models that combine TDA-derived features with deep learning techniques might, then, be particularly effective in enhancing the prediction of cultural cusps, ensuring that subtle but significant shifts in social behavior are accurately identified and anticipated.

Therefore, a feasible model inspired in Asimov's vision could potentially be employed by relating Persistent Homologies to the concepts designed by the behaviorist approach. In my research, I have identified three Persistent Homologies categories, such as Nuclear, Bipolar, and Multipolar Constellations (2024a), which could closely correlate with the behaviorist concepts of Macrobehavior, Macrocontingency, and Cultural Cusps.

Nuclear Constellations, for instance, represent tightly-knit clusters within social media networks that exhibit strong, cohesive interactions over time. These constellations might mirror the concept of Macrobehavior, where persistent interaction patterns within a group reinforce group cohesion and possibly worsen a scenario of political polarization – though such a statement would require further exploration. It must be disclaimed, however, that in the scope of my pilot research, these Nuclear Constellation indicated processes of Political Personalism. Regardless, by analyzing these Nuclear Constellations through TDA, it is possible to identify stable topological features that indicate cohesive group behavior, providing information into the dynamics of political factions and their long-term stability. Moreover, the concept of Macrocontingencies also directly applies to Nuclear Constellations, as they encapsulate the aggregated effects of numerous individual actions converging towards a central point of influence. In a Nuclear Constellation, the central cluster is a manifestation of these Macrocontingencies, where the collective behaviors of individuals coalesce around a pivotal topic or influential figure. This aggregation not only amplifies the impact of the central theme but also reinforces the behaviors and interactions within the group, creating a powerful feedback loop. The echo chamber effect observed in Nuclear Constellations further exemplifies Macrocontingencies at work. As users repeatedly engage with similar content and like-minded individuals, their behaviors are continually reinforced, leading to a more pronounced and cohesive central cluster. This concentration of interaction and shared behavior highlights the macro-level impact of individual actions aggregated within the social media landscape.

Nuclear Constellations are characterized by a dense, central cluster of data points indicating a concentrated area of interaction or discussion. This formation suggests significant activity or attention around a specific narrative stream – which was confirmed in the scope of the 2022 Brazilian elections on Twitter as Nuclear Constellations always showcased cohesion through political personalism (Rocha, 2023, 2024b). The k-Nearest-Neighbors (kNN, where **k = 3**) filtration is employed after a reduction of the point cloud (to enable analysis in limited hardware) to uncover the persistent homology within these constellations by examining the nearest neighbors and revealing the underlying topological structure, which is clearly nucleated. Below, each point represents an user, and the vertexes as their nearest neighbor connections.

**Figure 1** – A Nuclear Constellation identified through Twitter Interactions.

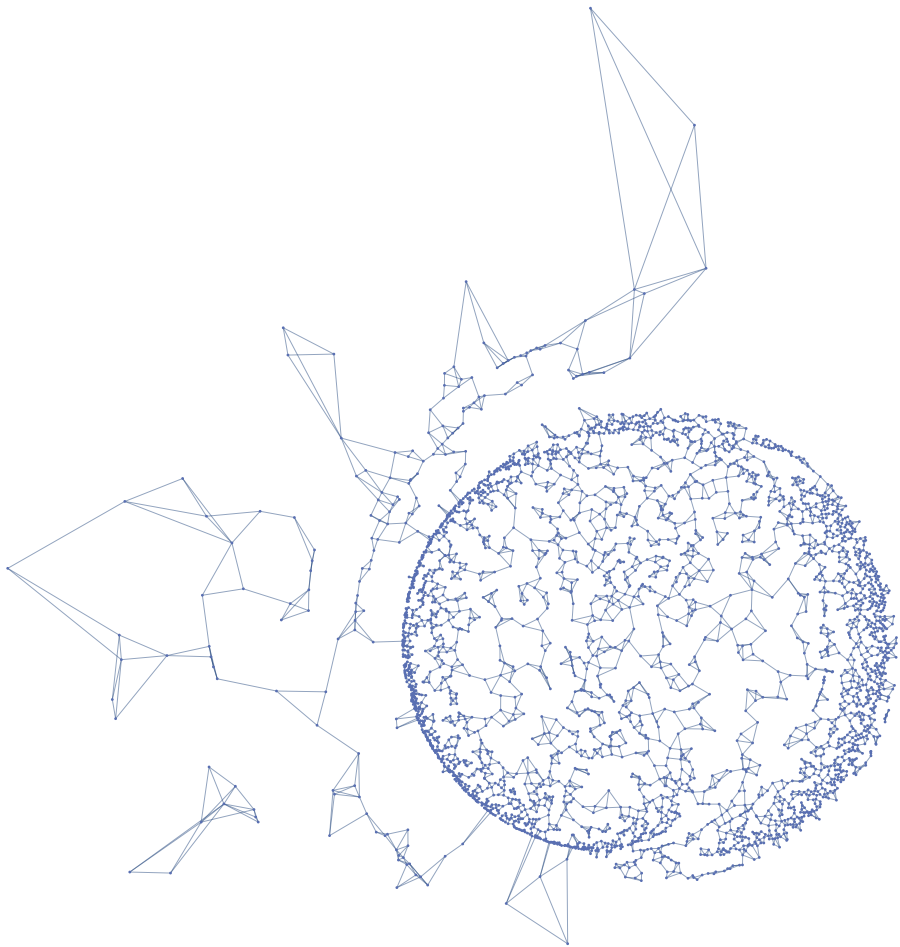

**Source:** Rocha, 2024a.

Cultural Cusps, as previously stated, represent pivotal moments where unique behavioral contingencies converge, leading to substantial sociocultural changes. In the context of Nuclear Constellations, these cusps can be seen as the moments when a particular topic, event, or influential figure suddenly garners widespread attention, resulting in the formation of a dense, central cluster of interactions. For instance, a viral event or a significant political announcement can act as a Cultural Cusp, around which a Nuclear Constellation forms, indicating a concentrated and cohesive response from the online community. By analyzing the timing and context of these clusters, researchers can identify the key moments that triggered significant shifts in public discourse and engagement, reflecting the impact of Cultural Cusps.

Nuclear Constellations can, overtime, evolve into Bipolar Constellations. In Bipolar Constellations, Cultural Cusps can be associated with events or issues that sharply divide public opinion, leading to the formation of two opposing clusters – and the gatekeepers involved in the narrative stream split. These cusps might include polarizing political events, contentious social issues, or influential media reports that provoke strong reactions from different segments of the population (Rocha, 2023). The emergence of Bipolar Constellations around these cusps highlights the points of divergence in societal views – or political personalism – and the subsequent polarization. By studying the formation and evolution of these Bipolar Constellations, it is possible to pinpoint the Cultural Cusps that catalyzed the division and analyze how these moments contribute to the broader landscape of political polarization.

Bipolar Constellations exhibit two dense clusters of data points, representing opposing areas of interaction or discussion. In the scope of the 2022 Brazilian Presidential Elections on Twitter, these constellations are indicative of polarized discourse, often prevalent in political discussions. kNN filtrations were also used to identify the nearest neighbors within each cluster, highlighting the separation and interaction patterns between the clusters. The distance between the centroids of these clusters provides insights into the degree of polarization, with kNN helping to reveal the distinct topological features and the minimal interactions between the opposing groups. Alike Figure 1, each point on this graph represents an user, while the lines here display the nearest neighbor connections (kNN) between users, with **k = 3**.

**Figure 2** – A Bipolar Constellation identified through Twitter Interactions.

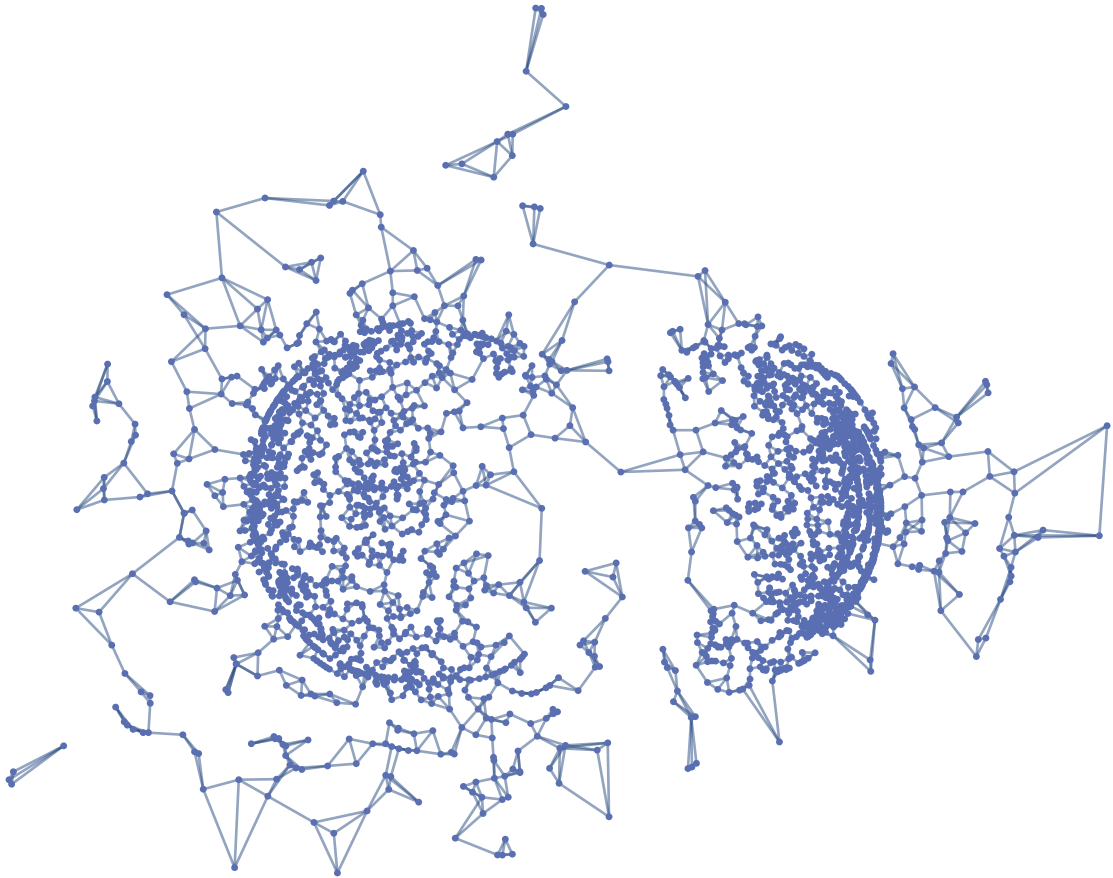

**Source:** Rocha, 2024a.

This comprehensive pilot approach to integrating AI with TDA in the analysis of social media data provides a potentially powerful framework for understanding societal trends. From there, and by identifying and modeling these key behaviorist concepts, researchers might be able to develop predictive models that capture the complexity and dynamism of social behaviors, moving closer to realizing the vision of Psychohistory. It is worth noting that further studies, particularly in regard to the temporal graphs, have to be employed to conclude significant remarks about the role of Multipolar Constellations.

For instance, during the 2022 Brazilian Presidential Elections, certain debates or controversies acted as Cultural Cusps, resulting in distinct Bipolar Constellations on Twitter. By employing k-Nearest-Neighbors (kNN) filtrations, it was possible identify the nearest neighbors within each cluster, revealing the separation and interaction patterns between the clusters, with the distance between the centroids of these clusters providing information as to the degree of polarization, with kNN helping to reveal the distinct topological features and the minimal interactions between the opposing groups.

Subsequent research will have to consider: Temporal Analysis, by using TDA to track the evolution of constellations over time, identifying the exact moments when clusters form or change in response to Cultural Cusps; Cluster Dynamics, to analyze the stability and transformation of clusters before, during, and after Cultural Cusps to better understand their long-term impact on social media networks and reveal how certain events consolidate or fragment public opinion over time; and finally, Behavioral Patterns that examine he individual behaviors and interactions that contribute to the formation of constellations around Cultural Cusps, highlighting the key drivers of sociocultural change, involving the investigation of influential actors, types of content shared, and the nature of interactions (e.g., supportive or contentious).

## Conclusions

This exploratory paper is but an early literature review aiming to bridge the visionary concept of Asimov's Psychohistory with contemporary computational methodologies, particularly TDA and AI. By delving into the intersections of social media analysis, behaviorist concepts, and advanced mathematical frameworks, I set the stage for a new paradigm in forecasting societal trends. Nevertheless, further research is necessary to investigate the relationship between behaviorist concepts and Persistent Homologies: this involves systematically tracking Metacontingencies, Macrobehaviors, Macrocontingencies, and Cultural Cusps to uncover the intricate patterns that underpin social dynamics in the scope of Social Media Content as understanding these relationships will enhance the predictive power of TDA and AI in analyzing large social media datasets. Additionally, further research will investigate the emerging field of Sociophysics, a field of science which uses mathematical tools derived from Physics to understand the behavior of crowds, as its inherent connection with TDA might reveal interesting tools for the furthering development of this research.

An essential avenue for future studies is the role of gatekeepers in social media networks. These influential actors shape narratives and filter echo chambers (Garimella, et. Al, 2018), making them crucial for identifying Cultural Cusps—key moments of sociocultural change which will likely showcase moments of cluster splitting, or reconfigurations of Nuclear Constellations into Bipolar, or Multipolar Constellations. By mapping the activities and influence patterns of gatekeepers, it is possible to realize the triggers and propagation of significant societal shifts. This focus on gatekeepers is particularly important for Persistent Homology Time Series Analysis (Rocha, Silva, Vicentini, 2024), as it showcases how loops (holes) might develop into Bipolar Constellations, creating new cultural cusps and clusters. Understanding these dynamics provides meaningful information into the evolution of societal trends, highlighting the role of key actors in the emergence of significant sociopolitical changes.

Social media platforms, with their vast data reflecting human interactions and behaviors, are indispensable for actualizing the concept of Psychohistory – and might, in fact, be the deciding factor that allowed fiction to become reality. They provide a unique lens through which the complexities of modern society can be observed, analyzed, and eventually predicted. As our computational capabilities continue to evolve, the vision of predicting large-scale social behaviors with precision comes closer to reality.

In summary, this paper highlights the potential of integrating TDA and AI techniques with social media data to advance our understanding of societal trends. By building on the legacy of Psychohistory, we aim to inspire further research that holistically incorporates historical, psychological, mathematical, and computational studies, paving the way for significant advancements in the field of Computational Social Sciences.